\pdfoutput=1

\documentclass[12pt,a4paper]{article}

\usepackage{ifthen} 
\newboolean{pdflatex}
\setboolean{pdflatex}{true} 

\newboolean{articletitles}
\setboolean{articletitles}{true} 

\newboolean{uprightparticles}
\setboolean{uprightparticles}{false} 

\newboolean{inbibliography}
\setboolean{inbibliography}{false} 

\def\paperauthors{LHCb collaboration} 
\def\paperasciititle{Observation of the doubly Cabibbo-suppressed Xicp->p phi decay} 
\def\papertitle{Observation of the doubly Cabibbo-suppressed decay \xicppphi} 
\def\paperkeywords{{High Energy Physics}, {LHCb}} 
\def\papercopyright{\the\year\ CERN for the benefit of the LHCb collaboration} 
\def\paperlicence{CC-BY-4.0 licence}
\def\paperlicenceurl{https://creativecommons.org/licenses/by/4.0/}

\usepackage[top=1in, bottom=1.25in, left=1in, right=1in]{geometry}

\usepackage{microtype}
\usepackage{lineno}  
\usepackage{xspace} 
\usepackage{caption} 

\usepackage{graphicx}  
\usepackage{color}
\usepackage{colortbl}
\graphicspath{{./}} 

\usepackage{amsmath} 
\usepackage{amssymb}
\usepackage{amsfonts}
\usepackage{upgreek} 

\newcommand*\patchAmsMathEnvironmentForLineno[1]{%
\expandafter\let\csname old#1\expandafter\endcsname\csname #1\endcsname
\expandafter\let\csname oldend#1\expandafter\endcsname\csname
end#1\endcsname
 \renewenvironment{#1}%
   {\linenomath\csname old#1\endcsname}%
   {\csname oldend#1\endcsname\endlinenomath}%
}
\newcommand*\patchBothAmsMathEnvironmentsForLineno[1]{%
  \patchAmsMathEnvironmentForLineno{#1}%
  \patchAmsMathEnvironmentForLineno{#1*}%
}
\AtBeginDocument{%
\patchBothAmsMathEnvironmentsForLineno{equation}%
\patchBothAmsMathEnvironmentsForLineno{align}%
\patchBothAmsMathEnvironmentsForLineno{flalign}%
\patchBothAmsMathEnvironmentsForLineno{alignat}%
\patchBothAmsMathEnvironmentsForLineno{gather}%
\patchBothAmsMathEnvironmentsForLineno{multline}%
\patchBothAmsMathEnvironmentsForLineno{eqnarray}%
}

\usepackage{hyperxmp}

\usepackage[pdftex,
            pdfauthor={\paperauthors},
            pdftitle={\paperasciititle},
            pdfkeywords={\paperkeywords},
            pdfcopyright={Copyright (C) \papercopyright},
            pdflicenseurl={\paperlicenceurl}]{hyperref}

\usepackage[all]{hypcap} 

\usepackage{xspace} 
\usepackage{upgreek}

\def\lhcb {\mbox{LHCb}\xspace}

\def\rich   {RICH\xspace}

\def\MagUp {\mbox{\em Mag\kern -0.05em Up}\xspace}

\ifthenelse{\boolean{uprightparticles}}%
{

 \def\Ppi         {\ensuremath{\uppi}\xspace}

 \def\Pphi        {\ensuremath{\upphi}\xspace}

 \def\PDelta      {\ensuremath{\Delta}\xspace}                 
 \def\PXi      {\ensuremath{\Xi}\xspace}                 
 \def\PLambda      {\ensuremath{\Lambda}\xspace}                 
 \def\PSigma      {\ensuremath{\Sigma}\xspace}                 
 \def\POmega      {\ensuremath{\Omega}\xspace}                 
 \def\PUpsilon      {\ensuremath{\Upsilon}\xspace}                 
 
 \def\PB      {\ensuremath{\mathrm{B}}\xspace}                 
                  
 \def\PD      {\ensuremath{\mathrm{D}}\xspace}

 \def\PK      {\ensuremath{\mathrm{K}}\xspace}

 \def\PW      {\ensuremath{\mathrm{W}}\xspace}

 \def\Pb      {\ensuremath{\mathrm{b}}\xspace}                 
 \def\Pc      {\ensuremath{\mathrm{c}}\xspace}                 
 \def\Pd      {\ensuremath{\mathrm{d}}\xspace}

 \def\Pi      {\ensuremath{\mathrm{i}}\xspace}

 \def\Pp      {\ensuremath{\mathrm{p}}\xspace}

 \def\Ps      {\ensuremath{\mathrm{s}}\xspace}                 
                  
 \def\Pu      {\ensuremath{\mathrm{u}}\xspace}

}
{

 \def\Ppi         {\ensuremath{\pi}\xspace}

 \def\Pphi        {\ensuremath{\phi}\xspace}

 \mathchardef\PDelta="7101
 \mathchardef\PXi="7104
 \mathchardef\PLambda="7103
 \mathchardef\PSigma="7106
 \mathchardef\POmega="710A
 \mathchardef\PUpsilon="7107
                  
 \def\PB      {\ensuremath{B}\xspace}                 
                  
 \def\PD      {\ensuremath{D}\xspace}

 \def\PK      {\ensuremath{K}\xspace}

 \def\PW      {\ensuremath{W}\xspace}

 \def\Pb      {\ensuremath{b}\xspace}                 
 \def\Pc      {\ensuremath{c}\xspace}                 
 \def\Pd      {\ensuremath{d}\xspace}

 \def\Pi      {\ensuremath{i}\xspace}

 \def\Pp      {\ensuremath{p}\xspace}

 \def\Ps      {\ensuremath{s}\xspace}                 
                  
 \def\Pu      {\ensuremath{u}\xspace}

}

\makeatletter
\ifcase \@ptsize \relax
  \newcommand{\miniscule}{\@setfontsize\miniscule{4}{5}}
\or
  \newcommand{\miniscule}{\@setfontsize\miniscule{5}{6}}
\or
  \newcommand{\miniscule}{\@setfontsize\miniscule{5}{6}}
\fi
\makeatother

\DeclareRobustCommand{\optbar}[1]{\shortstack{{\miniscule (\rule[.5ex]{1.25em}{.18mm})}
  \\ [-.7ex] $#1$}}

\def\Wp     {{\ensuremath{\PW^+}}\xspace}

\def\uquark    {{\ensuremath{\Pu}}\xspace}

\def\dquark    {{\ensuremath{\Pd}}\xspace}

\def\squark    {{\ensuremath{\Ps}}\xspace}
\def\squarkbar {{\ensuremath{\overline \squark}}\xspace}

\def\cquark    {{\ensuremath{\Pc}}\xspace}

\def\bquark    {{\ensuremath{\Pb}}\xspace}

\def\pion   {{\ensuremath{\Ppi}}\xspace}

\def\pip    {{\ensuremath{\pion^+}}\xspace}
\def\pim    {{\ensuremath{\pion^-}}\xspace}

\def\kaon    {{\ensuremath{\PK}}\xspace}
  \def\Kbar    {{\kern 0.2em\overline{\kern -0.2em \PK}{}}\xspace}

\def\KorKbar    {\kern 0.18em\optbar{\kern -0.18em K}{}\xspace}

\def\Kp      {{\ensuremath{\kaon^+}}\xspace}
\def\Km      {{\ensuremath{\kaon^-}}\xspace}

\newcommand{\phiz}{\ensuremath{\Pphi}\xspace}

  \def\Dbar    {{\kern 0.2em\overline{\kern -0.2em \PD}{}}\xspace}
\def\D       {{\ensuremath{\PD}}\xspace}

\def\DorDbar    {\kern 0.18em\optbar{\kern -0.18em D}{}\xspace}

\def\Dp      {{\ensuremath{\D^+}}\xspace}

\def\Dsp     {{\ensuremath{\D^+_\squark}}\xspace}

\def\Bbar    {{\ensuremath{\kern 0.18em\overline{\kern -0.18em \PB}{}}}\xspace}

\def\BorBbar    {\kern 0.18em\optbar{\kern -0.18em B}{}\xspace}

  \def\Y#1S{\ensuremath{\PUpsilon{(#1S)}}\xspace}

\def\proton      {{\ensuremath{\Pp}}\xspace}

\def\Xires       {{\ensuremath{\PXi}}\xspace}

\def\Lz          {{\ensuremath{\PLambda}}\xspace}
\def\Lbar        {{\ensuremath{\kern 0.1em\overline{\kern -0.1em\PLambda}}}\xspace}
\def\LorLbar    {\kern 0.18em\optbar{\kern -0.18em \PLambda}{}\xspace}

\def\Lc      {{\ensuremath{\Lz^+_\cquark}}\xspace}

\def\Xicp    {{\ensuremath{\Xires^+_\cquark}}\xspace}

\def\BF         {{\ensuremath{\mathcal{B}}}\xspace}

\newcommand{\decay}[2]{\ensuremath{#1\!\to #2}\xspace}         

\def\to                 {\ensuremath{\rightarrow}\xspace}

\def\Vud  {{\ensuremath{V_{\uquark\dquark}}}\xspace}
\def\Vcd  {{\ensuremath{V_{\cquark\dquark}}}\xspace}

\def\Vus  {{\ensuremath{V_{\uquark\squark}}}\xspace}
\def\Vcs  {{\ensuremath{V_{\cquark\squark}}}\xspace}

\def\Vcds  {{\ensuremath{V_{\cquark\dquark}^\ast}}\xspace}

\def\AT#1     {\ensuremath{A_{\mathrm{T}}^{#1}}\xspace}           

\def\C#1      {\ensuremath{\mathcal{C}_{#1}}\xspace}                       
\def\Cp#1     {\ensuremath{\mathcal{C}_{#1}^{'}}\xspace}                    
\def\Ceff#1   {\ensuremath{\mathcal{C}_{#1}^{\mathrm{(eff)}}}\xspace}        
\def\Cpeff#1  {\ensuremath{\mathcal{C}_{#1}^{'\mathrm{(eff)}}}\xspace}       
\def\Ope#1    {\ensuremath{\mathcal{O}_{#1}}\xspace}                       
\def\Opep#1   {\ensuremath{\mathcal{O}_{#1}^{'}}\xspace}                    

\newcommand{\tev}{\ifthenelse{\boolean{inbibliography}}{\ensuremath{~T\kern -0.05em eV}}{\ensuremath{\mathrm{\,Te\kern -0.1em V}}}\xspace}
\newcommand{\gev}{\ensuremath{\mathrm{\,Ge\kern -0.1em V}}\xspace}
\newcommand{\mev}{\ensuremath{\mathrm{\,Me\kern -0.1em V}}\xspace}
\newcommand{\kev}{\ensuremath{\mathrm{\,ke\kern -0.1em V}}\xspace}
\newcommand{\ev}{\ensuremath{\mathrm{\,e\kern -0.1em V}}\xspace}
\newcommand{\gevc}{\ensuremath{{\mathrm{\,Ge\kern -0.1em V\!/}c}}\xspace}
\newcommand{\mevc}{\ensuremath{{\mathrm{\,Me\kern -0.1em V\!/}c}}\xspace}
\newcommand{\gevcc}{\ensuremath{{\mathrm{\,Ge\kern -0.1em V\!/}c^2}}\xspace}
\newcommand{\gevgevcccc}{\ensuremath{{\mathrm{\,Ge\kern -0.1em V^2\!/}c^4}}\xspace}
\newcommand{\mevcc}{\ensuremath{{\mathrm{\,Me\kern -0.1em V\!/}c^2}}\xspace}

\def\mum  {\ensuremath{{\,\upmu\mathrm{m}}}\xspace}

\def\invfb   {\ensuremath{\mbox{\,fb}^{-1}}\xspace}

\def\ps   {\ensuremath{{\mathrm{ \,ps}}}\xspace}

\newcommand{\chisqip}{\ensuremath{\chi^2_{\text{IP}}}\xspace}

\def\gsim{{~\raise.15em\hbox{$>$}\kern-.85em
          \lower.35em\hbox{$\sim$}~}\xspace}
\def\lsim{{~\raise.15em\hbox{$<$}\kern-.85em
          \lower.35em\hbox{$\sim$}~}\xspace}

\def\sPlot{\mbox{\em sPlot}\xspace}

\def\ptot       {\mbox{$p$}\xspace}
\def\pt         {\mbox{$p_{\mathrm{ T}}$}\xspace}

\def\evtgen     {\mbox{\textsc{EvtGen}}\xspace}

\def\geant      {\mbox{\textsc{Geant4}}\xspace}

\def\photos     {\mbox{\textsc{Photos}}\xspace}

\def\pythia     {\mbox{\textsc{Pythia}}\xspace}

\def\tell1  {TELL1\xspace}
\def\ukl1   {UKL1\xspace}

\newcommand{\ie}{\mbox{\itshape i.e.}\xspace}

\newcommand{\cf}{\mbox{\itshape cf.}\xspace}

\usepackage[colorinlistoftodos,textsize=scriptsize]{todonotes}

\usepackage{cite} 
\usepackage{mciteplus}
\def\cf      {CF\xspace}
\def\scs     {SCS\xspace}
\def\dcs     {DCS\xspace}
\def\pid     {PID\xspace}

\def\mVud  {{\ensuremath{\vert V_{\uquark\dquark}\vert }}\xspace}
\def\mVcd  {{\ensuremath{\vert V_{\cquark\dquark}\vert }}\xspace}

\def\mVus  {{\ensuremath{\vert V_{\uquark\squark}\vert }}\xspace}
\def\mVcs  {{\ensuremath{\vert V_{\cquark\squark}\vert }}\xspace}

\def\xicppphi   {\decay{\Xicp}{\proton\phiz}}

\def\xicppkk    {\decay{\Xicp}{\proton\Km\Kp}}
\def\xicppkh    {\decay{\Xicp}{\proton\Km h^{+}}}
\def\xicppkpi   {\decay{\Xicp}{\proton\Km\pip}}
\def\phikk      {\decay{\phiz}{\Kp\Km}}

\def\lzppi      {\decay{\Lz}{\proton\pim}}

\def\lcppkk     {\decay{\Lc}{\proton\Km\Kp}}
\def\lcppkpi    {\decay{\Lc}{\proton\Km\pip}}
\def\lcppkh     {\decay{\Lc}{\proton\Km h^{+}}}
\def\lcppkpiDCS {\decay{\Lc}{\proton\Kp\pim}}

\def\pkk    {\ensuremath{\proton\Km\Kp}\xspace}

\def\kk     {\ensuremath{\Km\Kp}\xspace}

\def\pkpi   {\ensuremath{\proton\Km\pip}\xspace}
\def\pkh    {\ensuremath{\proton\Km h^{+}}\xspace}

\def\Mkk    {\ensuremath{M_{\Km\Kp}}\xspace}

\def\Mpkk   {\ensuremath{M_{\proton\Km\Kp}}\xspace}
\def\Mpkpi  {\ensuremath{M_{\proton\Km\pip}}\xspace}
\def\Mpkh   {\ensuremath{M_{\proton\Km h^{+}}}\xspace}

\usepackage{longtable} 
\usepackage{graphpap}
\usepackage{rotating}

\begin{document}

\renewcommand{\thefootnote}{\fnsymbol{footnote}}
\setcounter{footnote}{1}


\begin{titlepage}
\pagenumbering{roman}

\vspace*{-1.5cm}
\centerline{\large EUROPEAN ORGANIZATION FOR NUCLEAR RESEARCH (CERN)}
\vspace*{1.5cm}
\noindent
\begin{tabular*}{\linewidth}{lc@{\extracolsep{\fill}}r@{\extracolsep{0pt}}}
\ifthenelse{\boolean{pdflatex}}
{\vspace*{-1.5cm}\mbox{\!\!\!\includegraphics[width=.14\textwidth]{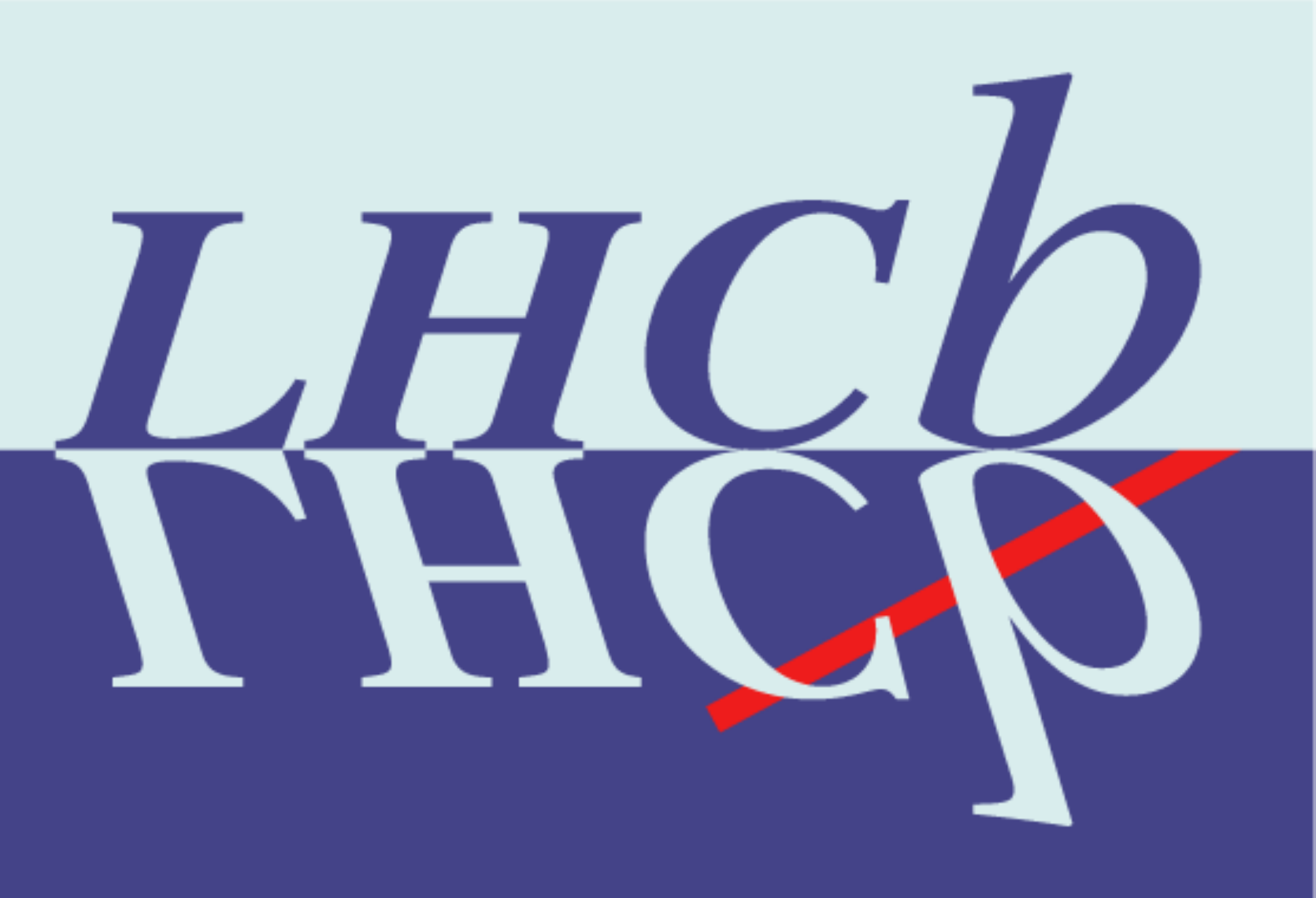}} & &}%
{\vspace*{-1.2cm}\mbox{\!\!\!\includegraphics[width=.12\textwidth]{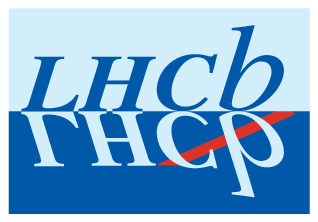}} & &}%
\\
 & & CERN-EP-2018-349 \\  
 & & LHCb-PAPER-2018-040 \\  
 & & January 18, 2019 \\
 & & \\
\end{tabular*}

\vspace*{4.0cm}

{\normalfont\bfseries\boldmath\huge
\begin{center}
  \papertitle 
\end{center}
}

\vspace*{2.0cm}

\begin{center}
\paperauthors\footnote{Authors are listed at the end of this paper.}
\end{center}

\vspace*{0.3cm}
This paper is dedicated to the memory of our friend and colleague Yury Shcheglov.
\vspace*{0.6cm}


\begin{abstract}
  \noindent
  The doubly Cabibbo-suppressed decay \xicppphi with \phikk is observed  for the first 
  time, with a statistical significance of more than 
  fifteen standard deviations.
  The data sample used in this analysis corresponds to an integrated luminosity of 
  2\invfb recorded with the \lhcb detector in \proton\proton collisions 
  at a centre-of-mass energy of 8\tev. The ratio of branching fractions between the 
  decay \xicppphi and the singly Cabibbo-suppressed decay \xicppkpi is measured to be
  \begin{equation*}
    \frac{\BF(\xicppphi)}{\BF(\xicppkpi)} = (19.8 \pm 0.7 \pm 0.9 \pm 0.2)
    \times 10^{-3},
  \end{equation*}
  where the first uncertainty is statistical, the second systematic and the third due to the knowledge of the $\phi\to\Kp\Km$ branching fraction.
\end{abstract}

\vspace*{0.8cm}

\begin{center}
  Submitted to JHEP
\end{center}


{\footnotesize 
\centerline{\copyright~\papercopyright. \href{\paperlicenceurl}{\paperlicence}.}}
\vspace*{2mm}

\end{titlepage}


\newpage
\setcounter{page}{2}
\mbox{~}

\cleardoublepage


\renewcommand{\thefootnote}{\arabic{footnote}}
\setcounter{footnote}{0}


\pagestyle{plain} 
\setcounter{page}{1}
\pagenumbering{arabic}


\section{Introduction}
\label{sec:Introduction}
The flavour structure of the weak interaction between quarks is described by the
Cabibbo-Kobayashi-Maskawa (CKM) matrix~\cite{Cabibbo:1963yz,*Kobayashi:1973fv}.
In particular, the tree-level decays of charmed particles depend on the matrix 
elements \Vud, \Vus, \Vcd and \Vcs. The hierarchy of the CKM matrix elements becomes 
evident using the approximate Wolfenstein parametrisation, which is based on the 
expansion in powers of the small parameter $\lambda\approx0.23$ with
\mbox{\mVud$\approx\,$\mVcs$\approx1-\lambda^{2}/2$} and 
\mVus$\approx\,$\mVcd$\approx\lambda$~\cite{PhysRevLett.51.1945,PDG2018}.
Tree-level decays depending on both \Vus and \Vcd matrix elements are known as doubly 
Cabibbo-suppressed (\dcs) decays. They have small branching fractions compared 
to the Cabibbo-favoured (\cf) and the singly Cabibbo-suppressed (\scs) 
decays~\cite{PhysRevD.55.7067}.  A systematic study of the relative contributions of 
\dcs and \cf diagrams to decays of charm baryons could shed light onto the role of the 
nonspectator quark, and in particular Pauli interference~\cite{Bianco:2003vb}. 
Such studies would be helpful for a better understanding of the lifetime hierarchy
of charm baryons~\cite{Bianco:2003vb,Blok:1991st,Cheng:2015rra,LHCb-PAPER-2018-028}.
So far only one \dcs charm-baryon decay, \lcppkpiDCS, has been 
observed~\cite{PhysRevLett.117.011801,LHCb-PAPER-2017-026}.

\begin{figure}[h]
  \setlength{\unitlength}{1mm}
  \centering
  \begin{picture}(100,40)
    \put( 20, 0){\includegraphics*[width=60mm,height=40mm]{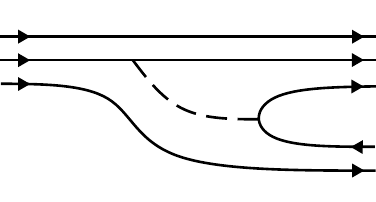}}
    \put( 6,27){\Large\Xicp}
    \put(17,32.2){\uquark}
    \put(17,27.2){\cquark}
    \put(17,22.2){\squark}
    \put(87,7){\Large\phiz}
    \put(81,5){\squark}
    \put(81,10){\squarkbar}
    \put(87, 28){\Large\proton}
    \put(81, 32.2){\uquark}
    \put(81, 27.2){\dquark}
    \put(81, 22.2){\uquark}
    \put(56,12){\Vus}
    \put(36,23.5){\Vcds}
    \put(49,20){\Wp}
  \end{picture}
  \caption {\small
     Tree quark diagram for the \xicppphi decay.
  }
  \label{fig:feynman}
\end{figure}

This article reports the first observation of the \dcs decay \xicppphi 
with \phikk, hereafter referred to as the signal decay 
channel.\footnote{The inclusion of charge-conjugated processes is implied 
throughout this article.} The leading-order 
diagram for the \xicppphi decay is shown in Fig.~\ref{fig:feynman}. 
The branching fraction of the signal decay channel is measured relative to the 
branching fraction of the \scs decay channel \xicppkpi, 
\begin{equation}
    R_{p\phi}\equiv\frac{\BF(\xicppphi)}{\BF(\xicppkpi)}\,.
\end{equation}
The measurement is based on a data sample of \proton\proton collisions collected in
2012 with the \lhcb detector at the centre-of-mass energy of 8\tev , corresponding 
to an integrated luminosity of 2\invfb.

\section{Detector and simulation}
\label{sec:Detector}

The \lhcb detector~\cite{Alves:2008zz,LHCb-DP-2014-002} is a single-arm forward
spectrometer covering the \mbox{pseudorapidity} range $2<\eta <5$,
designed for the study of particles containing \bquark or \cquark
quarks. The detector includes a high-precision tracking system
consisting of a silicon-strip vertex detector surrounding the $pp$
interaction region~\cite{LHCb-DP-2014-001}, a large-area silicon-strip detector 
located upstream of a dipole magnet with a bending power of about
$4{\mathrm{\,Tm}}$, and three stations of silicon-strip detectors and straw
drift tubes~\cite{LHCb-DP-2013-003} placed downstream of the magnet.
The tracking system provides a measurement of the momentum, \ptot, of charged 
particles with a relative uncertainty that varies from 0.5\% at low momentum to 1.0\% 
at 200\gevc. The minimum distance of a track to a primary vertex (PV), the impact 
parameter (IP), is measured with a resolution of $(15+29/\pt)\mum$,
where \pt is the component of the momentum transverse to the beam, in\,\gevc.
Different types of charged hadrons are distinguished using information
from two ring-imaging Cherenkov detectors~\cite{LHCb-DP-2012-003}. 
Photons, electrons, and hadrons are identified by a system consisting of
scintillating-pad and preshower detectors, an electromagnetic and a hadronic 
calorimeter. Muons are identified by a
system composed of alternating layers of iron and multiwire
proportional chambers~\cite{LHCb-DP-2012-002}.
The online event selection is performed by a trigger~\cite{LHCb-DP-2012-004}, 
which consists of a hardware stage, based on information from the calorimeter 
and the muon systems, followed by a software stage, which applies a full event
reconstruction.

At the hardware trigger stage, the events are required to have a muon with high \pt 
or a hadron, photon or electron with high transverse energy in the calorimeters. 
The software trigger requires a two-, three- or four-track secondary vertex with 
a significant displacement from any primary $pp$ interaction vertex. At least 
one charged particle must have a transverse momentum $\pt > 1.6\gevc$ and be 
inconsistent with originating from any PV.

Simulation is used to evaluate detection efficiencies for the signal and the 
normalisation decay channels. In the simulation, $pp$ collisions are generated using 
\pythia~\cite{Sjostrand:2006za,*Sjostrand:2007gs} with the specific \lhcb
configuration~\cite{LHCb-PROC-2010-056}. Decays of hadronic particles are 
described by \evtgen~\cite{Lange:2001uf}, in which the final-state radiation is 
generated using \photos~\cite{Golonka:2005pn}. The interaction of the generated 
particles with the detector and its response are implemented using the \geant
toolkit~\cite{Allison:2006ve, *Agostinelli:2002hh} as described in
Ref.~\cite{LHCb-PROC-2011-006}.
\section{Selection of candidates}
\label{sec:Selection}

The candidates for the \xicppkh decays, where $h^+=\{\pip,\Kp\}$, are formed using three 
charged tracks with $\pt>250\mevc$. Hadrons used for the reconstruction of the \Xicp 
baryons should not be produced at the PV. Only pions, protons, and kaons with an impact 
parameter \chisqip in excess of 9 with respect to all reconstructed PVs are taken into 
consideration for subsequent analysis. The \chisqip quantity is calculated as the 
difference in $\chi^2$ of the PV fit with and without the particle in question. The 
momenta of the reconstructed final-state particles are required to be in the range 
3.2 -- 150\gevc for the mesons, and in the range 10 -- 100\gevc for the proton. 
The reconstructed tracks must pass particle-identification (PID) requirements based on 
information from the \rich detectors, the calorimeter, and the muon 
stations~\cite{LHCb-PUB-2016-021}. The \pid requirements are loose for mesons and much 
tighter for protons, to suppress \pip and \Kp misidentified as protons. The three tracks 
must form a common vertex. The selected \Xicp candidates must have the rapidity ($y$) 
and transverse momentum $2.0<y<4.5$ and $4<\pt<16\gevc$. 

Additional requirements are introduced to suppress the contribution from \Dp and 
\Dsp decays with pions or kaons misidentified as protons. Such background manifests
itself as narrow peaking structures in the mass spectrum of the three hadrons
if the mass hypothesis for the track identified as a proton is changed to a pion
or kaon. Candidates with a mass within $\pm10$\mevcc (approximately $\pm2.5\sigma$) 
of the known values are rejected.

The average number of visible interactions per beam-crossing is 1.7~\cite{LHCb-DP-2014-002}.
The candidate is associated to the PV with the smallest value of \chisqip.
In order to evaluate the candidate \Xicp decay time and the two-body masses 
for the particles in the final state, a constrained fit is performed,
requiring the \Xicp candidate to have originated from its associated PV and 
have a mass equal to its known value~\cite{Hulsbergen:2005pu}.
The proper decay time is required to be between 0.55 and 1.5\ps to reduce the fraction
of baryons coming from $b$-hadron decays. The $b$-hadron component is also 
suppressed by the requirement on the \chisqip value of the reconstructed 
baryon to be less than 32. The masses of the \pkh combinations are calculated 
without the mass constraint. They are required to be in the range 
\mbox{$2.42$ to $2.51\gevcc$} for the \Xicp candidates.

In the offline selection, trigger objects are associated with reconstructed 
particles~\cite{LHCb-DP-2012-004}. Selection requirements can therefore be made 
on the trigger selection itself and on whether the decision was due to the 
signal decay candidate (Trigger On Signal, TOS category), or to other particles 
produced in the $pp$ collision (Trigger Independent of Signal, TIS category) 
or to a combination of both. The selected candidates must belong to the TIS 
category of the hardware-trigger and to the TOS category of the two levels of 
the software-trigger. 

Only \xicppkk candidates from the \phikk region, \ie candidates with a \kk mass 
(\Mkk) less than 1.07\gevcc, are used. A very small fraction of \xicppphi 
events leaks into the $\Mkk>1.07\gevcc$ region. In the $R_{p\phi}$
measurement this effect is taken into account using the distribution observed 
in simulated events. Figures~\ref{fig:pphi} (left) and \ref{fig:pkpi} 
show the mass distribution of the selected candidates for the \xicppkk and
\xicppkpi decay channels, respectively. Clear peaks can be seen in both 
distributions. The studies of the underlying background events suggest 
no peaking contributions for the signal and normalisation decay channels.

In parallel to \Xicp selections, samples of \lcppkh decays are also selected. 
The candidates for the \Lc decays are used to calibrate resolutions 
and trigger efficiencies and to perform other cross-checks.
\section{Fit model and yields of signal and normalisation candidates}
\label{sec:Fit}

\begin{figure}[tb]
  \setlength{\unitlength}{1mm}
  \centering
  \begin{picture}(150,60)
    \put( 0,0){\includegraphics*[width=75mm,height=60mm]{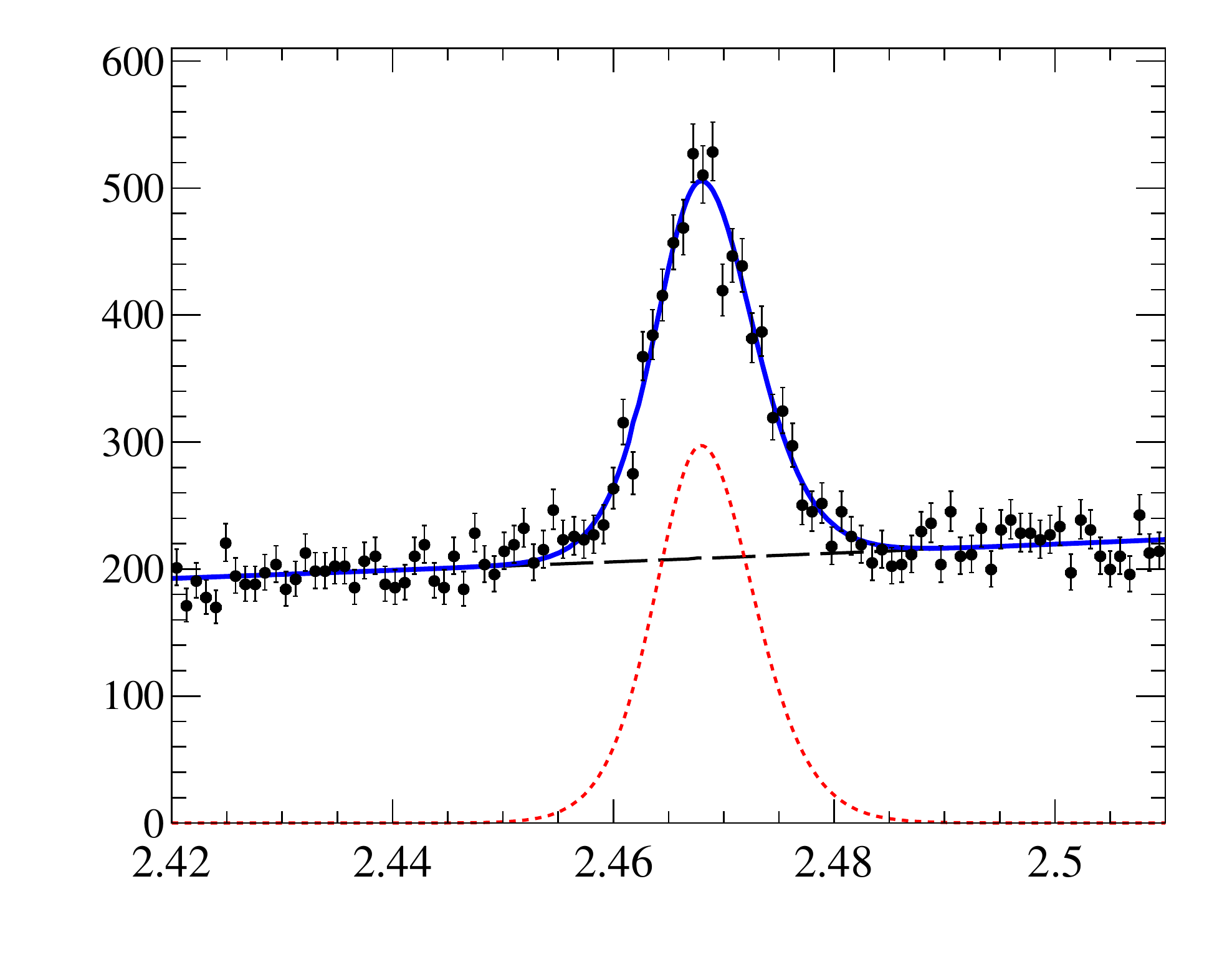}}
    \put(75,0){\includegraphics*[width=75mm,height=60mm]{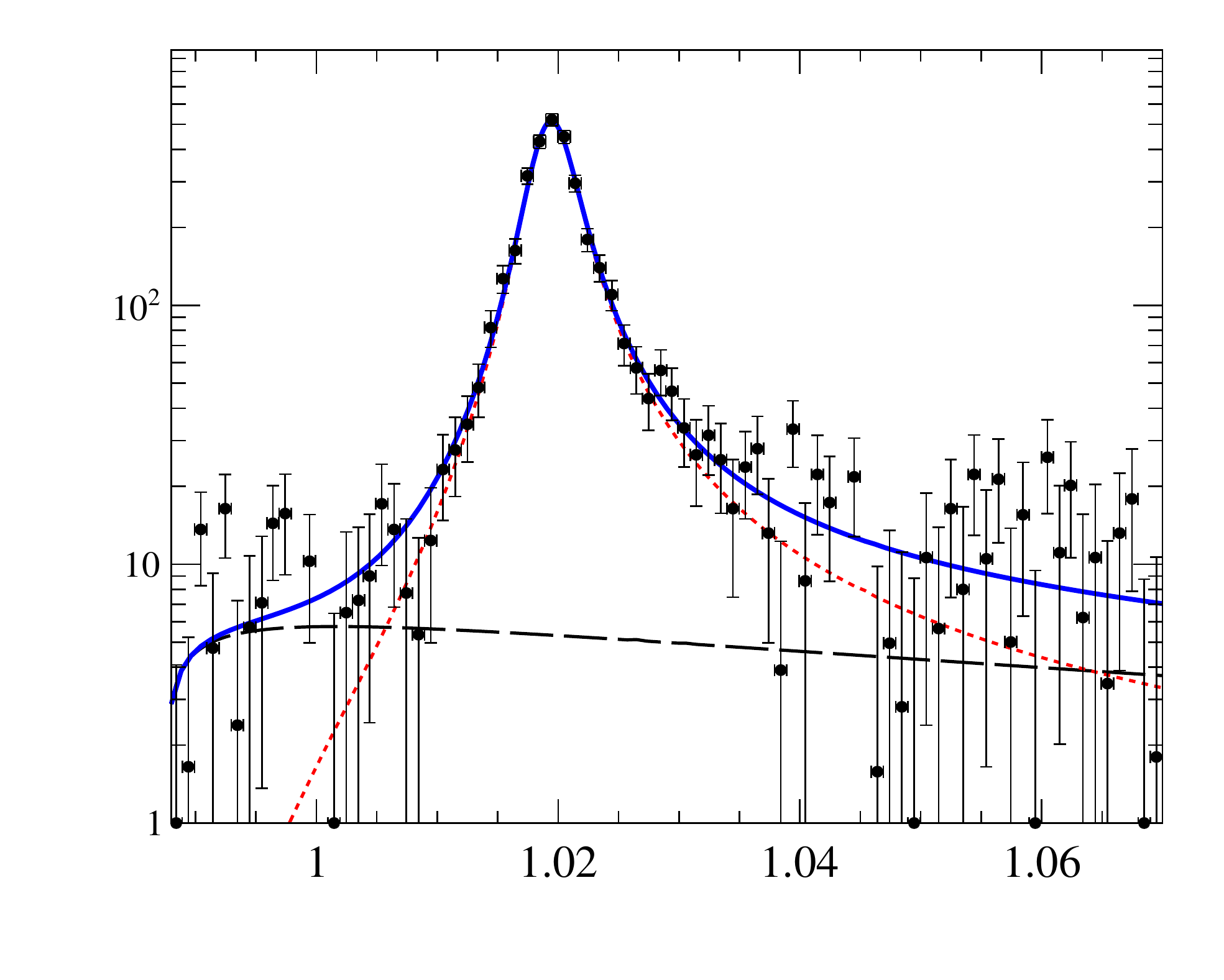}} 
    \put(  1,16){\begin{sideways}Candidates/$(1\mevcc)$\end{sideways}}
    \put( 77,16){\begin{sideways}Candidates/$(1\mevcc)$\end{sideways}}
    \put( 36,1){\Mpkk [\gevcc]}
    \put(113,1){\Mkk [\gevcc]}
    \put( 15,47){\lhcb}
    \put(120,47){\lhcb}
  \end{picture}
  \caption {\small
            (Left) Fit results for the \xicppkk decay. The candidates are selected in 
            the \phiz meson region, \ie with the requirement of $\Mkk<1.07\gevcc$.
            The red dotted line corresponds to 
            the signal component, the black dashed line reflects the background 
            distribution, and the blue solid line is their sum.
            (Right) Background subtracted \kk mass distribution for the \xicppkk decay.
            The red dotted line shows the \xicppphi
            contribution, the black dashed line represents the non-\phiz contribution,
            and the solid blue line is the total fit function.
  }
  \label{fig:pphi}
\end{figure}

\begin{figure}[tb]
  \setlength{\unitlength}{1mm}
  \centering
  \begin{picture}(150,60)
    \put( 40, 0){\includegraphics*[width=75mm,height=60mm]{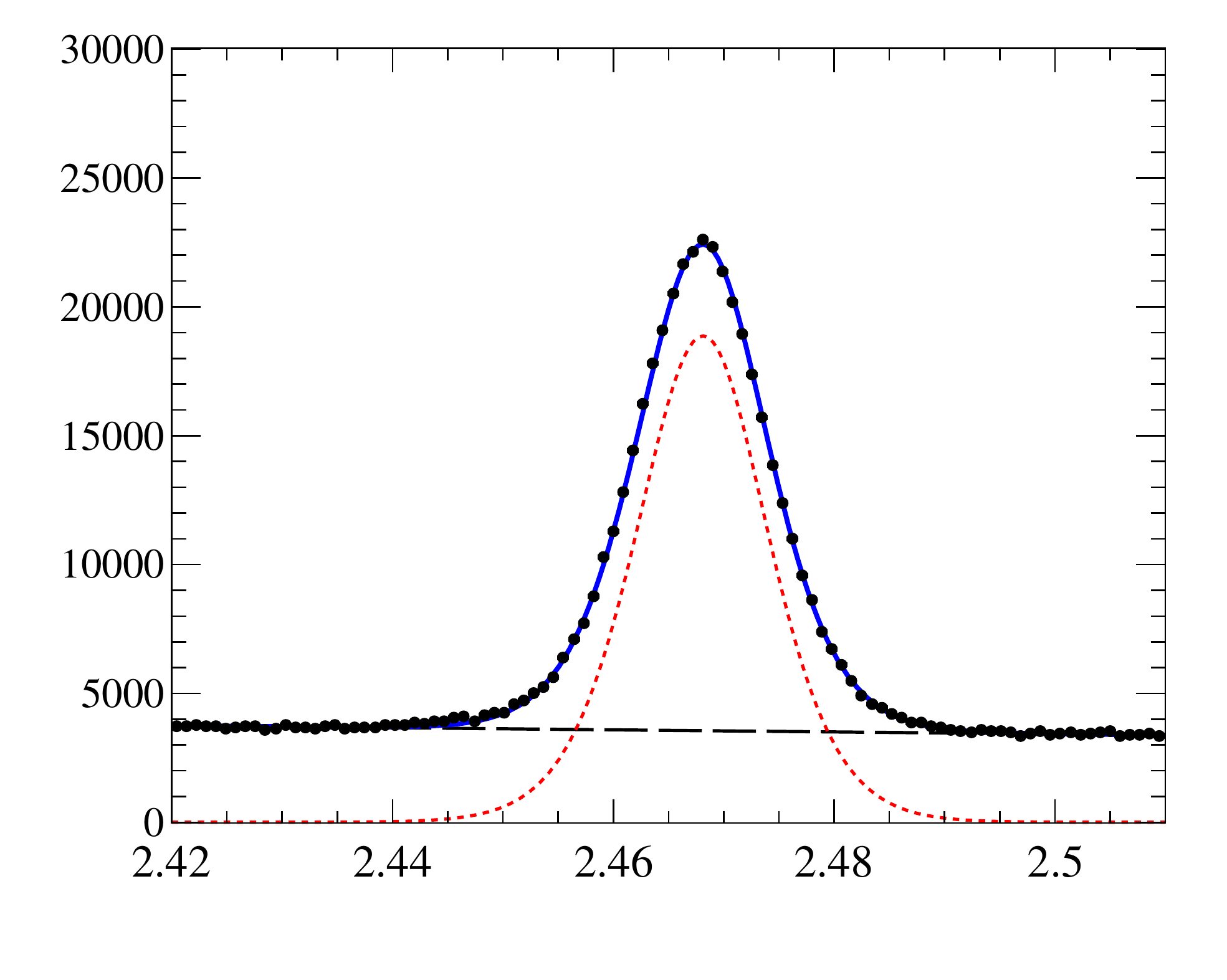}}
    \put( 38,16){\begin{sideways}Candidates/$(1\mevcc)$\end{sideways}}
    \put( 78,1){\Mpkpi [\gevcc]}
    \put( 57,47){\lhcb}
  \end{picture}
  \caption {\small
            Fit results for the \xicppkpi decay. The red dotted line corresponds to 
            the signal component, the black dashed line reflects the background 
            distribution and the blue solid line is their sum.
  }
  \label{fig:pkpi}
\end{figure}

The yields of the selected \xicppkh decays are determined from 
unbinned extended maximum-likelihood fits to the corresponding \pkk
or \pkpi mass spectra. The probability density function consists 
of a Gaussian core and exponential tails. The following distribution is used
as the \Xicp model: 
\begin{equation}
f_{\Xicp}(x,\beta)\propto\exp\big\{\beta^{2}-\sqrt{ \beta^4+ x^2\beta^2 }\big\}, \ \ \ 
x  = \frac{M-\mu}{\sigma( 1 + \epsilon \kappa) },
\label{eq:Apolonious2}
\end{equation}
where $M$ is the candidate mass, $\mu$ is the peak position, $\sigma$ reflects 
the core-peak width, $\kappa$ is an asymmetry parameter, and $\beta$ characterises 
the exponential tails~\cite{OstapGitHub}. The value of $\epsilon$ is $-1$ for 
$M\leq\mu$ and $+1$ for $M>\mu$. The parameter $\beta$ is fixed in the fit of the
\xicppkk mass distribution to the value obtained from the fits of the 
normalisation and of the \lcppkk decay channels. The background is modelled 
by an exponential function. The results of the fits for the \xicppkk and \xicppkpi 
decay channels are presented in Figs.~\ref{fig:pphi} and \ref{fig:pkpi}, 
respectively. The yields are $N_{pKK} = 3790\pm120$ for the \xicppkk decay
channel and $N_{pK\pi} = (324.7\pm0.8)\times10^{3}$ for the normalisation
decay channel.

To separate the \phiz and non-\phiz contributions to the signal decay channel, 
the background subtracted \kk mass distribution is analysed. The subtraction 
is done using the \sPlot technique~\cite{Pivk:2004ty}. The \Mkk observable is 
evaluated with the \Xicp mass constraint and is almost independent from the \Mpkk 
discriminating variable. The effect of the correlation is small and is taken into 
account in the systematic uncertainty of the measurement.

The fraction of the \phiz contribution ($f_{\phi}$) in the selected 
\xicppkk candidates is determined by a binned nonextended maximum-likelihood 
fit to the  \Mkk spectrum. A $P$-wave relativistic Breit--Wigner distribution with
Blatt--Weisskopf form factor~\cite{PhysRevD.5.624} is used to describe 
the \phikk lineshape. The barrier radius is set to 3.5~GeV$^{-1}$ in natural units.
This distribution is convolved with a Gaussian function to model the 
experimental resolution. The parameters of the resolution function are fixed
using the \lcppkk sample. For the non-\phiz contribution, the Flatt\'e 
parameterisation~\cite{FLATTE1976224} is used in the form
\begin{equation}
 f_{{\rm non}\text{-}\phiz}\propto\left\{m^{2}_{0}-\Mkk^{2} -im_{0}\left(g_{1}\rho_{\pi\pi}+g_{2}\rho_{KK}\right)\right\}^{-2},
 \label{eq:Flatte}
\end{equation}
where $m_0$ refers to the mass of the $f_{0}(980)$ resonance, $g_{1}$ 
and $g_2$ are coupling constants, and $\rho_{\pi\pi}$ and $\rho_{KK}$ 
are the Lorentz-invariant phase-space factors. The term $g_{2}\rho_{KK}$
accounts for the opening of the kaon threshold. The values 
$m_{0}g_{1}=0.165\pm0.018\gev^2$ and $g_{2}/g_{1} = 4.21\pm0.33$ have been 
determined by the BES collaboration~\cite{ABLIKIM2005243}. The choice of the 
Flatt\'e parametrisation is suggested by the \kk mass distribution in the 
\lcppkk data sample. The \phiz contribution dominates in the \kk mass spectrum 
with a measured fraction $f_{\phi} = (90.0 \pm 2.7)\%$. The reported statistical 
uncertainty of the $f_{\phi}$ parameter is determined by a set of the 
pseudoexperiments, in which toy samples are generated according to result 
obtained for the alternative two-dimensional (\Mpkk vs. \Mkk) model 
described below. 

As a cross-check of the result obtained with the \sPlot approach, an extended 
two-dimensional likelihood fit to the \Mpkk and \Mkk distributions is performed.
Four two-dimensional terms are considered. The \Mpkk dependency for the \phiz and 
non-\phiz terms for the \Xicp decay component are described by Eq.~\ref{eq:Apolonious2}.
Two additional \phiz and non-\phiz terms are introduced for the \Mpkk background 
description. These terms are independent linear distributions in the \Mpkk spectrum.
A second-order polynomial is used to describe the \kk mass distribution of the
non-\Xicp non-\phiz background. The results of the two-dimensional fit are in agreement 
with the \sPlot-based procedure.

The statistical significance of the observation of the \xicppphi 
decay is estimated using Wilks' theorem~\cite{Wilks:1938dza} and is well
above $15\sigma$. The fit to the \Mkk distribution results in an evidence of a 
non-\phiz contribution to the \dcs \xicppkk decay. A statistical significance of 
$3.9\sigma$ is obtained under the assumption of normal distributions for the 
uncertainties.

\section{Efficiencies and branching fractions ratio}
\label{sec:Efficiencies}

The total detection efficiencies for both the signal and the normalisation 
decays can be factorised as
\begin{equation}
\label{eq:eff}
\epsilon_{\rm total} = \epsilon_{\rm acc}\times\epsilon_{\rm rec\&sel|acc}\times\epsilon_{\rm software|rec\&sel}\times\epsilon_{\rm hardware|software}\times\epsilon_{\rm PID},
\end{equation}
where $\epsilon_{\rm acc}$ denotes the geometrical acceptance of the 
\lhcb detector, $\epsilon_{\rm rec\&sel|acc}$ corresponds to the efficiency 
of reconstruction and selection of the candidates within the geometrical 
acceptance, $\epsilon_{\rm hardware|software}$ and $\epsilon_{\rm software|rec\&sel}$ 
are the trigger efficiencies for the selected candidates of the hardware 
and software levels, respectively, and $\epsilon_{\rm PID}$ is the \pid 
efficiency. Since the hardware trigger level accepts events 
independently of the reconstructed candidates, \ie the events belong to the 
TIS category, the efficiency 
$\epsilon_{\rm hardware|software}$ is assumed to cancel in the ratio of the 
signal and normalisation efficiencies. All other efficiencies except 
$\epsilon_{\rm PID}$ are determined from simulation. The simulated sample of 
\xicppkk events with the intermediate \phiz resonance is used to determine
efficiencies for the signal decay channel. The simulated sample for the \xicppkpi 
decay was produced according to a phase-space distribution. It is corrected to 
reproduce the Dalitz plot distribution observed with data. An additional correction
is introduced for both simulated samples to account for the difference in the 
tracking efficiencies between data and simulation~\cite{LHCb-DP-2013-002}. 

The \pid efficiencies for the hadrons are determined from large 
samples of protons, kaons, and pions~\cite{LHCb-PUB-2016-021}.
These samples are binned in momentum and pseudorapidity of the hadron, as well
as in the charged particle multiplicity of the event. The \pid efficiency for 
the \Xicp candidates are determined on an event-by-event basis. The weights 
for each candidate are taken from the calibration histograms using trilinear 
interpolation. The efficiency $\epsilon_{\rm PID}$ is determined as the ratio 
of \Xicp yields obtained from maximum-likelihood fits of the \Mpkh distributions 
from the weighted and unweighted samples.

The ratio between the total efficiencies of the signal and the normalisation 
decay channels is determined in bins of \pt and $y$ of the \Xicp baryon. 
This procedure accounts for kinematic features of the \Xicp production, 
which could be poorly modelled in the simulation. Averaged over the $(\pt,y)$
bins this ratio is determined to be $(91.1\pm3.6)\%$, including systematic
uncertainties.

To reduce the effect of the dependence of the efficiency on the \Xicp 
kinematics, the mass fits are repeated in seven nonoverlapping $(\pt,y)$ 
bins, which cover the \lhcb fiducial volume. The fit procedure is the 
same as described above, except that the $\sigma$ parameter of the signal 
distribution in Eq.~\ref{eq:Apolonious2} is fixed to the value of the 
normalisation decay channel, scaled by a factor obtained from a fit to 
the \lcppkk and \lcppkpi mass distributions in the same $(\pt,y)$ bins. 
The ratios of the yields of the signal and normalisation decay channels 
are corrected by the ratios of the total efficiencies. The branching 
fraction ratios are evaluated for each 
$(\pt,y)$ bin as
\begin{equation}
  R_{p\phi} = \frac{ N_{pKK}f_{\phi}}{\BF(\phikk)}\times
  \frac{1}{N_{pK\pi}}\times
  \frac{\epsilon_{\rm total}^{pK\pi}}{\epsilon_{\rm total}^{p\phi}}.
\end{equation}
The known value of $\BF(\phikk)=0.492\pm0.005$ is used~\cite{PDG2018}.
The weighted average of the branching fraction ratios evaluated for the 
$(\pt,y)$ bins is $R_{p\phi} = (19.8 \pm 0.7) \times 10 ^{-3}$,
where the uncertainty reflects the statistical uncertainty of the \Xicp 
yields and $f_{\phi}$. The alternative two-dimensional fitting procedure 
gives $R_{p\phi} = (19.8\pm0.8)\times10^{-3}$, which is in excellent
agreement with the result determined using the \sPlot technique.

\section{Systematic uncertainties}
\label{sec:Systematic}

The list of systematic uncertainties for the measured ratio $R_{p\phi}$
is presented in Table~\ref{tab:syst}. The total uncertainty is
obtained as the quadratic sum of all contributions. 

\begin{table}[t]
\caption{Systematic uncertainties relative to the central value of the ratio $R_{p\phi}$.}
\label{tab:syst}
\centering
\begin{tabular}{lcc}
\hline
 Source                                   & Uncertainty (\%) \\
\hline
Signal fit model                          & 0.5      \\
Background fit model                      & 0.5      \\
\sPlot-related uncertainty                & 1.0      \\
Trigger efficiency                        & 3.0      \\
\pid efficiency                           & 2.2      \\
Tracking                                  & 1.0      \\
(\pt,$y$) binning                         & 1.3      \\
Size of simulation sample                 & 0.7      \\
Selection requirements                    & 0.8      \\
\hline
Total                                     & 4.4      \\
\hline
\end{tabular}
\end{table}

In order to estimate the systematic uncertainties for the yields of the 
\xicppkk and the normalisation decay channels, various hypotheses are tested 
for the description of the signal and background shapes. When the signal 
parameterisations in the \Mpkk and \Mpkpi spectra are changed to a modified 
Novosibirsk function~\cite{Lees:2011gw}, no significant 
deviation from the nominal fit model is found. The change of the function for 
the non-\phiz component to a two-body phase space model in the fit to the \Mkk 
distribution leads to a systematic uncertainty of 0.5\%, which is considered 
as the signal fit-model uncertainty.

The background-model parameterisation is tested by replacing of polynomial function
with a product of polynomial and exponential functions. The uncertainty related to 
the \sPlot method is studied with two samples of 500 pseudoexperiments each, in 
which the samples are generated according to the \Mpkk--\Mkk model described in
Sec.~\ref{sec:Fit}. In one set of pseudoexperiments the effect of the residual
correlation between \Mpkk and \Mkk is introduced. The systematic uncertainty of 
the \sPlot technique is assigned from the deviations of the results of these
tests from the nominal ones.

The cancellation of the hardware-trigger efficiencies in the ratio of 
the signal and the normalisation decay channels is studied with the \Lc 
control samples. A technique based on the partial overlap 
of the TIS and TOS subsamples~\cite{LHCb-DP-2012-004} is used to 
evaluate hardware efficiencies for the \lcppkh decay channels.
The data are consistent with the hypothesis  of equal hardware-trigger 
efficiencies for the signal and normalisation decay channels.
The precision achieved by means of these studies, limited by 
the statistics in the overlap between the TIS and TOS subsamples, 
is used as a systematic uncertainty for the hardware-trigger 
efficiency ratio. 

For the software-trigger, the systematic uncertainty is assessed 
using simulation. The large variation of software-trigger requirements
demonstrates the stability of the ratio of software-trigger efficiencies
for the signal and normalisation decay channels at the 1\% to 2\% level. 
The overall systematic uncertainty for both hardware- and software-trigger
efficiencies is dominated by the former and is reported in 
Table~\ref{tab:syst}.

The main source of uncertainty of the \pid efficiency is related to the 
difference between results obtained with different calibration samples
for the protons. The \lcppkpi sample is used as default in the analysis,
while results obtained with the \lzppi calibration sample are used 
to assign a systematic uncertainty. For determination of \pid 
efficiencies the calibration samples are binned according to proton, pion, 
or kaon kinematics. The associated systematic uncertainty is studied by 
comparing the results with different binning and interpolation schemes.
The uncertainty related to the finite size of the calibration samples is
considered to be fully correlated between the signal and normalisation decay
channels and to cancel in the ratio.

The dominant uncertainty on the tracking efficiency correction  arises from 
the different track reconstruction efficiency for kaons and pions due to 
different hadronic cross-sections with the detector material. Half of the
\Km\pip detection asymmetry measured by \lhcb~\cite{LHCb-PAPER-2014-013}
is assigned as systematic uncertainty. Another source of uncertainty due to 
tracking efficiency is related to the binning of the tracking correction 
histogram. The difference between the results using interpolated and binned
values of the efficiency is assigned as systematic uncertainty.

The uncertainty due to the selected $(\pt,y)$-bins to determine $R_{p\phiz}$ 
is obtained from studies carried out with an alternative binning. There is an
uncertainty of 0.7\% from the size of the simulation sample. The obtained 
value of $R_{p\phiz}$ is stable within 0.8\% against a variation of selection
requirements. This value is taken as the uncertainty due to the selection 
requirements. The uncertainty related to the Dalitz plot correction procedure
applied to the simulated sample is estimated by a variation of the $R_{p\phi}$
ratio obtained with different binnings of the histogram used for this 
correction. This uncertainty is found to be small with respect to other 
sources of uncertainty.

\section{Conclusions}
\label{sec:Conclusion}

The first observation of the \dcs \xicppphi decay is presented,
using \proton\proton collision data collected with the \lhcb
detector at a centre-of-mass energy of 8\tev, corresponding to 
an integrated luminosity of 2\invfb. The ratio of the branching fractions 
with respect to the \scs \xicppkpi decay channel is measured to be
\begin{equation*}
  R_{p\phi} = (19.8 \pm 0.7 \pm 0.9 \pm 0.2)\times 10^{-3},
\end{equation*}
where the first uncertainty is statistical, the second systematic and 
the third due to the knowledge of the $\phiz\to\Kp\Km$ branching fraction.
An evidence of the $3.5\sigma$, including systematic uncertainties, 
for a non-\phiz contribution to the \dcs \xicppkk decay is also found.

\section*{Acknowledgements}
%
%

\noindent We express our gratitude to our colleagues in the CERN
accelerator departments for the excellent performance of the LHC. We
thank the technical and administrative staff at the LHCb
institutes.
We acknowledge support from CERN and from the national agencies:
CAPES, CNPq, FAPERJ and FINEP (Brazil); 
MOST and NSFC (China); 
CNRS/IN2P3 (France); 
BMBF, DFG and MPG (Germany); 
INFN (Italy); 
NWO (Netherlands); 
MNiSW and NCN (Poland); 
MEN/IFA (Romania); 
MSHE (Russia); 
MinECo (Spain); 
SNSF and SER (Switzerland); 
NASU (Ukraine); 
STFC (United Kingdom); 
NSF (USA).
We acknowledge the computing resources that are provided by CERN, IN2P3
(France), KIT and DESY (Germany), INFN (Italy), SURF (Netherlands),
PIC (Spain), GridPP (United Kingdom), RRCKI and Yandex
LLC (Russia), CSCS (Switzerland), IFIN-HH (Romania), CBPF (Brazil),
PL-GRID (Poland) and OSC (USA).
We are indebted to the communities behind the multiple open-source
software packages on which we depend.
Individual groups or members have received support from
AvH Foundation (Germany);
EPLANET, Marie Sk\l{}odowska-Curie Actions and ERC (European Union);
ANR, Labex P2IO and OCEVU, and R\'{e}gion Auvergne-Rh\^{o}ne-Alpes (France);
Key Research Program of Frontier Sciences of CAS, CAS PIFI, and the Thousand Talents Program (China);
RFBR, RSF and Yandex LLC (Russia);
GVA, XuntaGal and GENCAT (Spain);
the Royal Society
and the Leverhulme Trust (United Kingdom);
Laboratory Directed Research and Development program of LANL (USA).

\addcontentsline{toc}{section}{References}
\setboolean{inbibliography}{true}
\bibliographystyle{LHCb}
\bibliography{standard,LHCb-PAPER,LHCb-CONF,LHCb-DP,LHCb-TDR,local}

\newpage

\newpage
\centerline{\large\bf LHCb collaboration}
\begin{flushleft}
\small
R.~Aaij$^{28}$,
C.~Abell{\'a}n~Beteta$^{46}$,
B.~Adeva$^{43}$,
M.~Adinolfi$^{50}$,
C.A.~Aidala$^{78}$,
Z.~Ajaltouni$^{6}$,
S.~Akar$^{61}$,
P.~Albicocco$^{19}$,
J.~Albrecht$^{11}$,
F.~Alessio$^{44}$
% ...

{\footnotesize \it
% ...
$ ^{6}$Universit{\'e} Clermont Auvergne, CNRS/IN2P3, LPC, Clermont-Ferrand, France\\
% ...
$ ^{11}$Fakult{\"a}t Physik, Technische Universit{\"a}t Dortmund, Dortmund, Germany\\
% ...
$ ^{19}$INFN Laboratori Nazionali di Frascati, Frascati, Italy\\
% ...
$ ^{28}$Nikhef National Institute for Subatomic Physics, Amsterdam, Netherlands\\
% ...
$ ^{43}$Instituto Galego de F{\'\i}sica de Altas Enerx{\'\i}as (IGFAE), Universidade de Santiago de Compostela, Santiago de Compostela, Spain\\
$ ^{44}$European Organization for Nuclear Research (CERN), Geneva, Switzerland\\
% ...
$ ^{46}$Physik-Institut, Universit{\"a}t Z{\"u}rich, Z{\"u}rich, Switzerland\\
% ...
$ ^{50}$H.H. Wills Physics Laboratory, University of Bristol, Bristol, United Kingdom\\
% ...
$ ^{61}$University of Cincinnati, Cincinnati, OH, United States\\
% ...
$ ^{63}$Syracuse University, Syracuse, NY, United States\\
% ...
$ ^{78}$University of Michigan, Ann Arbor, United States, associated to $^{63}$\\
% ...
}
\end{flushleft}
\end{document}